\documentclass[pra,twocolumn,showpacs,superscriptaddress]{revtex4-1}
\usepackage{graphicx}
\usepackage{bm}
\usepackage{amssymb}
\usepackage{color}
\usepackage{amsmath}
\usepackage{amstext}
\usepackage{latexsym}
\usepackage[usenames,dvipsnames]{xcolor}
\usepackage[colorlinks=true,citecolor=purple,linkcolor=RubineRed,urlcolor=purple]{hyperref}
\usepackage{bbold}
\usepackage{float}
\usepackage{enumerate}
\usepackage{braket}

\definecolor{pink}{rgb}{1,0.078,0.57}

\definecolor{green}{rgb}{0,0.7,0.9}

\newcommand{\beq}{\begin{equation}}
\newcommand{\eeq}{\end{equation}}
\newcommand{\beqa}{\begin{eqnarray}}
\newcommand{\eeqa}{\end{eqnarray}}

\begin{document}

\title{Entanglement Dynamics with a Stochastic Non-Hermitian Hamiltonian away from Exceptional Points
}

\author{Hamid Sakhouf}
\email{hamidsakhouf@gmail.com}
\affiliation{Beijing Computational Science Research Center, Beijing 100084, China}
\author{Peng Xue\href{https://orcid.org/0000-0002-4272-2883}{\includegraphics[scale=0.05]{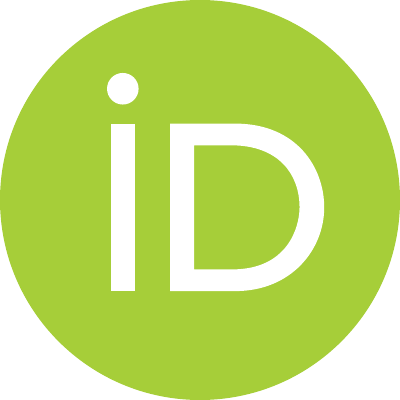}}}
\email{gnep.eux@gmail.com}
\affiliation{Beijing Computational Science Research Center, Beijing 100084, China}
\affiliation{Key Laboratory of Quantum Materials and Devices of Ministry of Education, School of Physics, Southeast University, Nanjing 211189, China.}


\begin{abstract}

Although non-Hermitian dynamics near exceptional points (EPs) provide a route to accelerated entanglement generation, entanglement can also be generated far from EPs at comparable or even higher rates. However, the behavior of such entanglement in open systems remains largely unexplored, rendering it highly susceptible to environmental noise. Here, we study entanglement dynamics in two coupled qubits described by a non-Hermitian Hamiltonian, where the dissipative rates are subject to stochastic noise. We focus on the regime away from EPs. Our approach demonstrates that, even in the presence of classical noise, rich dynamical control can be achieved, enabling highly efficient entanglement generation with a timescale that is significantly shorter than that of both Hermitian systems and EP-based non-Hermitian protocols. Additionally, the timescale is independent of the number of qubits, highlighting favorable scalability for multipartite entanglement generation and facilitating integration into future photonic quantum processors. 

\end{abstract}

 \maketitle
 \textit{Introduction---}
Any realistic description of quantum dynamics must account for the inevitable interaction between quantum systems and their surrounding environment. Open quantum system theory provides the framework to model these environmental effects, with parity-time ($\mathcal{PT}$) symmetric  non-Hermitian and stochastic Hamiltonians emerging as powerful tools to access rich, non-trivial quantum dynamics, particularly in the context of entanglement manipulation under noisy conditions. Non-Hermitian systems are often investigated via EPs—where both eigenenergies and eigenstates coalesce—as a means of enhancing entanglement dynamics. EP-enhanced entanglement has been theoretically demonstrated in systems of two or more weakly coupled non-Hermitian qubits
 \cite{1, 2, 3, 4, 5}, and experimentally validated in non-unitary quantum walk platforms  with  photonic systems  \cite{Tang,6,7}. Separately, non-Hermitian acceleration of entanglement generation—enabled by a stronger dissipation rate of one qubit in coupled swap qubits without quantum jumps—has also been proposed \cite{8}. Moreover, conditional  entanglement amplification can be achieved via non-Hermitian superradiant dynamics, which provides a new pathway to generating highly entangled macroscopic quantum states \cite{9}.

In the context of quantum simulation, stochastic systems typically arise from environmental interactions, where coupling to noise in large-scale open quantum systems introduces dissipative dynamics  \cite{10,11,12,Chenu,13}. Such systems are also relevant to the study of quantum information scrambling \cite{14}. These scenarios can be effectively captured using the Monte Carlo wave function method \cite{15,16,17,18,19,20}. Although the stochastic nature of quantum measurements inherently hinders deterministic quantum state preparation in open quantum systems, feedback-based unitary control operations have emerged as a powerful solution. Such approaches enable deterministic remote entanglement between qubits  \cite{21,22,23,24,25,Martin,26}, and have been experimentally demonstrated \cite{27,28,29,30}. Furthermore, the use of stochastic effects in conjunction with two-qubit interactions to stabilize entanglement has also been theoretically demonstrated under realistic experimental conditions \cite{31}. 

Recent work has advanced beyond these frameworks by investigating quantum dynamics using non-Hermitian Hamiltonians subject to classical noise, with a focus on noise effects in the Hamiltonian’s dissipative part as described by the antidephasing master equation \cite{32,33}. This approach enables rich dynamical control, supporting a broader diversity of steady states and facilitating quantum state purification. To our knowledge, the effect of stochastic noise on the dissipation term and its influence on entanglement dynamics remains unstudied. This is critical for understanding entanglement evolution and enhancing measurement accuracy in open quantum systems.

In the present work, we study the entanglement dynamics of two non-Hermitian qubits subject to classical noise, focusing on strong dissipation in one qubit far from the EPs and on the effects of stochastic noise in the system’s dissipative sector. Unlike existing schemes, where quantum-jump processes typically suppress entanglement rapidly \cite{Gal}, our protocol demonstrates excellent practical performance. We observe a characteristic population pattern among the basis states that yields a maximally entangled state on a timescale nearly an order of magnitude shorter than that of EP‑based non‑Hermitian protocols. Furthermore, we characterize the evolution of concurrence far from the EPs in the presence of classical noise, and discuss its relation to the corresponding fidelity dynamics.  In addition, extending the protocol to more than two qubits preserves the same entanglement generation timescale, leading to qubit-number-independent dynamics for this characteristic time.  Experimentally, the proposed scheme  can be readily implemented across various quantum platforms, including superconducting qubit processors and  photonic systems.

\begin{figure}[t]
	\centering
	\includegraphics[width = \linewidth]{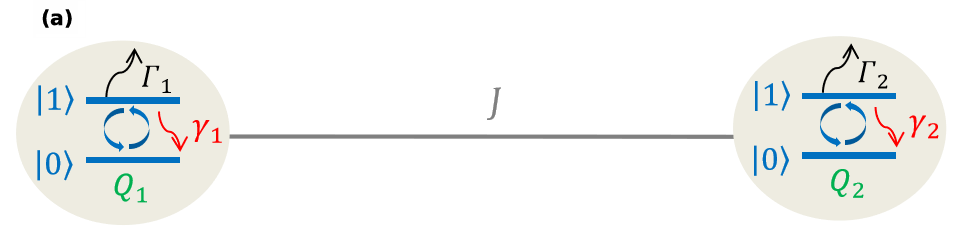}
	\hfill
	\includegraphics[width=\linewidth]{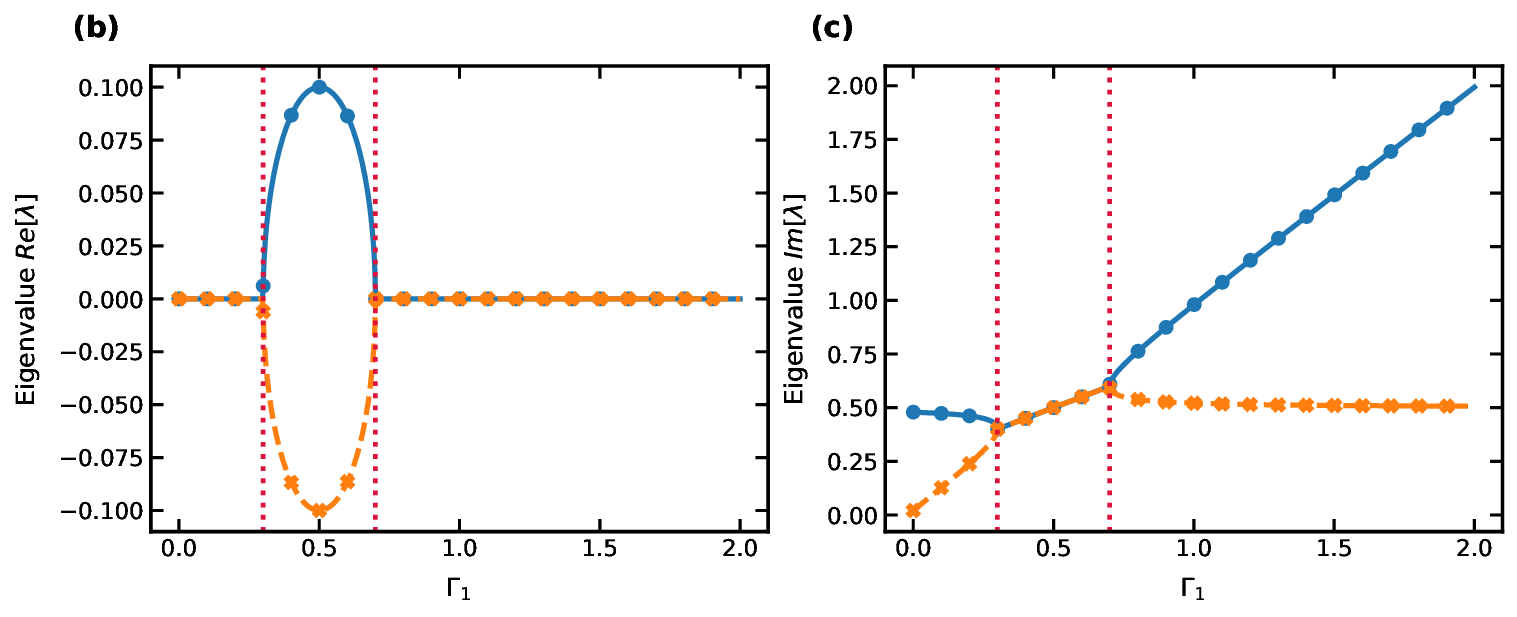}
	\caption{(a) Schematic of two non-Hermitian qubits, ${Q_1}$ and ${Q_1}$, coupled to each other with coupling strength $J$. The upper level of each qubit has a dissipation rate ${\Gamma _j}\left( {j = 1,2} \right)$ and a decay rate ${\gamma _j}\left( {j = 1,2} \right)$. (b,c) Real (b) and imaginary (c) parts of the eigenvalues as a function of ${\Gamma _1}$, for $J=0.1$ and ${\Gamma _2}=0.5$. }
	\label{Fig1}
\end{figure}

\emph{Model and dynamics under stochastic fluctuations---}
We consider a two-qubit non-Hermitian system, depicted in Fig. \ref{Fig1}(a), which is described by the Hamiltonian

\begin{equation}\label{Eq1}
	H = {H_{{\mathop{\rm int}} }} - i\sum\limits_{j = 1}^2 {{\Gamma _j}{{\left| 1 \right\rangle }_j}\left\langle 1 \right|} ,
\end{equation}
where ${{\Gamma _j}}$ are  dissipative rates and  ${H_{{\mathop{\rm int}} }} = J\left( {\sigma _1^ + \sigma _2^ -  + \sigma _1^ - \sigma _2^ + } \right)$ is  the  interaction  Hamiltonian with  coupling strength $J$ between the two qubits and $\sigma _j^ \mp $ are the lowering and  raising  operators of qubit $j$.
In  the  subspace $\left\{ {\left| {01} \right\rangle ,\left| {10} \right\rangle } \right\}$, the real and imaginary parts of the Hamiltonian eigenvalues are shown in Fig. \ref{Fig1}(b,c), with EPs occurring at ${\Gamma _1} = {\Gamma _2} \pm 2J$. Fixing  ${\Gamma _2}=0.5$ and varying ${\Gamma _1}$, the system exhibits two EPs located at ${\Gamma _1}=0.3$ and ${\Gamma _1}=0.7$.

In what follows, the Hermitian Hamiltonian ${H_{{\mathop{\rm int}} }}$
is subject to classical noise that couples to the anti-Hermitian operator $iL_j$ [see Fig. \ref{Fig1}(a)], resulting in the following  Hamiltonian
\begin{align} \label{Eq2} 
   \tilde H = {H_{{\mathop{\rm int}} }} - i\sum\limits_{J = 1}^2 {\left( {1 + \sqrt {2{\gamma _j}} {\xi _j}\left( t \right)} \right)} {L_j},
\end{align}
where the Hermitian operator ${L_j} = {\Gamma _j}{\left| 1 \right\rangle _j}\left\langle 1 \right|\left( {j = 1,2} \right)$, and  ${\gamma _j}$ are the decay rates, while ${\xi _j}\left( t \right)$ is  the white noise, satisfying $\left\langle {{\xi _j}\left( t \right)} \right\rangle  = 0$ and  $\left\langle {{\xi _j}\left( t \right){\xi _j}\left( {t'} \right)} \right\rangle  = \delta \left( {t - t'} \right)$. Here $\left\langle  \bullet  \right\rangle $ denotes
the classical stochastic average.
The system is then described by the
following the stochastic   master equation (SME) \cite{32}
\begin{align} \notag
   d\rho  =& \left( { - i\left[ {{H_{{\mathop{\rm int}} }},\rho } \right] - \left\{ {L_j,\rho } \right\} + {\gamma _j}\left\{ {L_j,\left\{ {L_j,\rho } \right\}} \right\}} \right)dt \\
   \label{Eq3}
   &- \sqrt {2{\gamma _j}} \left\{ {L_j,\rho } \right\}d{W_j}\left( t \right),
\end{align}
where $d{W_j}\left( t \right) = {\xi _j}\left( t \right)dt$ is  the Wiener process \cite{19,34,35}. For simplicity, we rewrite the  SME in Eq. (\ref{Eq3}) as
\begin{align}\notag 
   d\rho  =& \left[ { - i\left( {H\rho  - \rho {H^ + }} \right) + \frac{1}{2}\left\{ {{L_{\phi j}},\left\{ {{L_{\phi j}},\rho } \right\}} \right\}} \right]dt \\
   \label{Eq4}
   &- \left\{ {{L_{\phi j}},\rho } \right\}d{W_j}\left( t \right).
\end{align}
Here  ${L_{\phi j}} = \sqrt {{\gamma _{\phi j}}} {\left| 1 \right\rangle _j}\left\langle 1 \right|$ represent  the quantum
jump operators with the dephasing rate
 ${\gamma _{{\phi _j}}} = 2{\gamma _j}\Gamma _j^2$,  and  $H$ is the non-Hermitian Hamiltonian given in Eq. (\ref{Eq1}). In addition, we define the associated  stochastic Schrödinger equation (SSE) \cite{Martin,36,Gao,Kampen}  as follows
\begin{align}\notag 
	d\left| {\psi \left( t \right)} \right\rangle  =& \left[ { - iH + \frac{1}{2}L_{\phi j}^ + {L_{\phi j}}} \right]\left| {\psi \left( t \right)} \right\rangle dt\\
	\label{Eq5}
	 &- {L_{\phi j}}\left| {\psi \left( t \right)} \right\rangle d{W_j}\left( t \right).
\end{align}
 The dynamics of Eq. (\ref{Eq5}) are solved numerically, enforcing normalization to ensure trace preservation and focusing on the effect of quantum jumps  in the stochastic term (see Appendix \ref{app4} for the other term).

The two-qubit evolution in stochastic open quantum systems can be solved from the Hamiltonian in Eq. (\ref{Eq2}) via the SSE in Eq. (\ref{Eq5}). We use the populations of the two-qubit basis states and the concurrence to determine the timescale of the maximally entangled state, where the equality of the two populations in the subspace $\left\{ {\left| {01} \right\rangle ,\left| {10} \right\rangle } \right\}$ and the concurrence being maximized coincide exactly with the evolution time corresponding to the maximally entangled state. The concurrence in  open quantum systems \cite{37} is  given  by ${\cal C} = \max \left\{ {0,{\tau _1} - {\tau _2} - {\tau _3} - {\tau _4}} \right\}$, where ${\tau _1},{\tau _2},{\tau _3}$ and  ${\tau _4}$ are   the eigenvalues of the Hermitian matrix $R = \sqrt {\sqrt \rho  \tilde \rho \sqrt \rho  } $ with $\tilde \rho  = \left( {{\sigma _y} \otimes {\sigma _y}} \right){\rho ^ * }\left( {{\sigma _y} \otimes {\sigma _y}} \right)$, where ${\rho ^ * }$ is the complex conjugate of  ${\rho  }$ and  ${\sigma _y}$ is the Pauli  matrix. For simplification, we assume in our calculations that the two non-Hermitian qubits have identical decay rates for their respective excited states, i.e., ${\gamma _1} = {\gamma _2} \equiv \gamma $.

\begin{figure}[t]
 	\centering
 	\includegraphics[width = \linewidth]{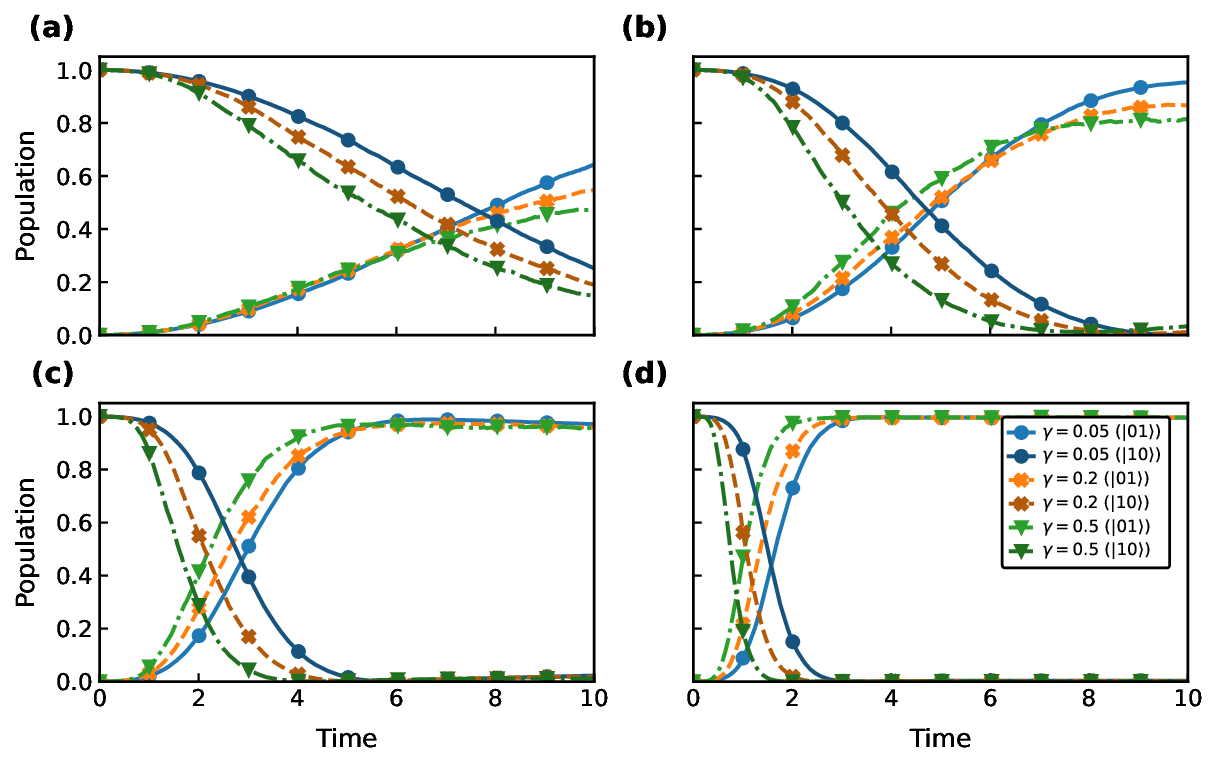}
 	\caption{ Populations of the two-qubit basis states as functions of the evolution time for different stochastic-noise strengths $\gamma$, $J=0.1$ and different values of the first-qubit dissipation rate $\Gamma_1$: (a) $\Gamma_1 = 0.5$, (b) $\Gamma_1 = 0.7$, (c) $\Gamma_1 = 1.1$, and (d) $\Gamma_1 = 2$. Calculations use the SSE in Eq. (\ref{Eq5}), and results are averaged over 1000 independent single trajectories with the initial state and  the dissipation rate of the second qubit are  given  by $\left| {\psi \left( 0 \right)} \right\rangle  = \left| {10} \right\rangle $ and $\Gamma_2 = 0.5$, respectively, for all calculations.  }
 	\label{Fig2}
 \end{figure}

\emph{Entanglement generation far from EPs with stochastic dissipative qubits---} 
 We first investigate the occupation probabilities of the $\left| {00} \right\rangle $, $\left| {01} \right\rangle $,  $\left| {10} \right\rangle $ and $\left| {01} \right\rangle $ basis states as a function of evolution time in our non-Hermitian system, accounting for stochastic fluctuations on the dissipative 
 rates for  different stochastic strengths ${\gamma}$, with the second qubit’s dissipation rate fixed at   ${\Gamma _2} = 0.5$, as shown in Fig. \ref{Fig2}. The dissipation rate of the first qubit ${\Gamma _1}$ is chosen across several values. Starting with ${\Gamma _1}=0.5$,  the system lies in the $\mathcal{PT}$-symmetry–unbroken phase regime [Fig. \ref{Fig2}(a)]. At ${\Gamma _1}=0.7$, the system reaches the second  EP [Fig. \ref{Fig2}(b)]. For larger values, ${\Gamma _1}=1.1$ and $2$, the system enters the $\mathcal{PT}$-symmetric-broken phase regime [Fig. \ref{Fig2}(c-d)]. Figure \ref{Fig2} shows that the populations of the  $\left| {01} \right\rangle $ and $\left| {10} \right\rangle $ states are exchanged during the coherent evolution and become equal at a well-defined interaction time, while the populations of the $\left| {00} \right\rangle $ and $\left| {11} \right\rangle $  states remain negligibly small throughout the evolution and are thus omitted from the plots. This behavior indicates the formation of a maximally entangled Bell state of the form ${{\left( {\left| {01} \right\rangle  \pm \left| {10} \right\rangle } \right)} \mathord{\left/
 		{\vphantom {{\left( {\left| {01} \right\rangle  \pm \left| {10} \right\rangle } \right)} {\sqrt 2 }}} \right.
 		\kern-\nulldelimiterspace} {\sqrt 2 }}$.
    
On the other hand, the reaction time varies inversely with the values of $\Gamma_1$; as we move further away from the EPs—that is, for ${\Gamma _1} \succ 0.7$ —the interaction time becomes shorter. Interestingly, the interaction time is nearly identical to that observed in the three-qubit case, indicating that it is independent of the number of qubits. This independence simplifies the experimental implementation. More details on the entanglement dynamics of the three-qubit case are provided in Appendix \ref{app3}.

\begin{figure}
    \centering
    \includegraphics[width = \linewidth]{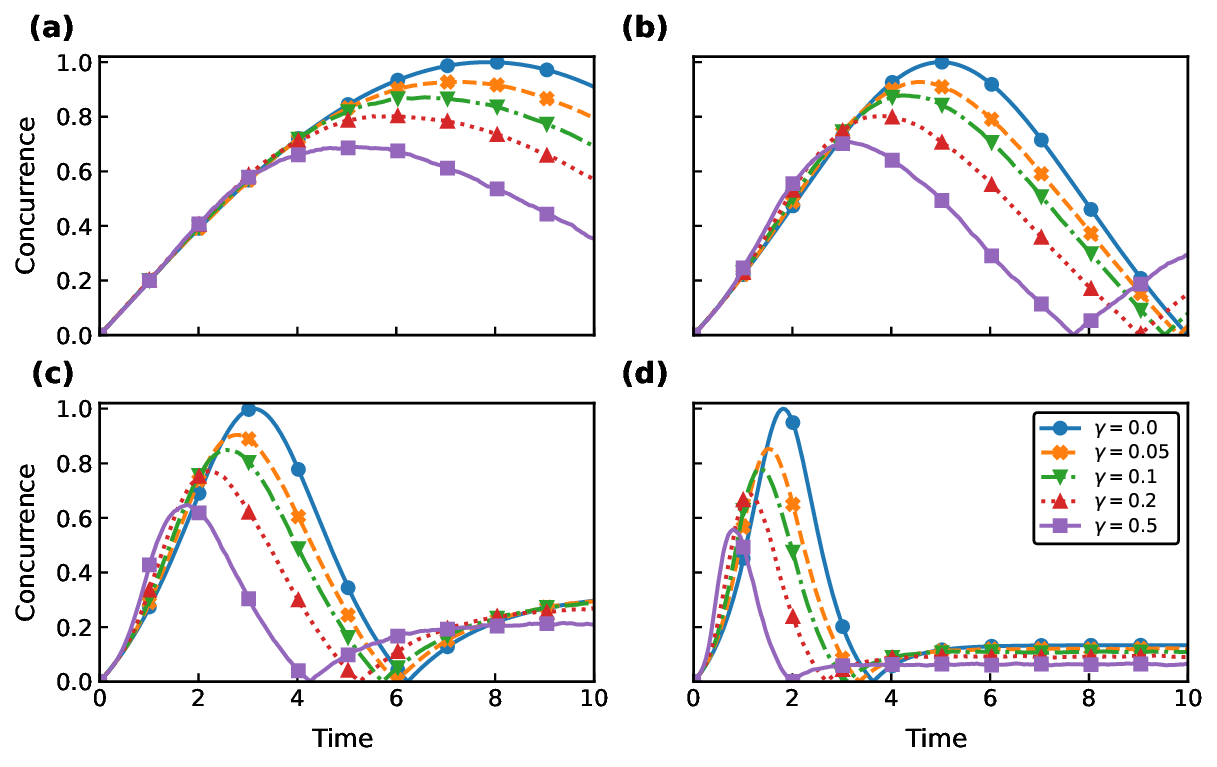}
    \caption{Mixed-state  concurrence as function  of evolution time  for  different  stochastic strenghs $\gamma$, with  ${\Gamma _2} = 0.5$, $J=0.1$ and for different values of ${\Gamma _1}$: (a) $\Gamma_1 = 0.5$, (b) $\Gamma_1 = 0.7$, (c) $\Gamma_1 = 1.1$, and (d) $\Gamma_1 = 2$. Calculations are performed based on the SSE given in Eq. (\ref{Eq5}), with the system initialized in the state $\left| {10} \right\rangle $  and results averaged over 1500 independent single trajectories. }
    \label{Fig3}
\end{figure}

We characterize the mixed-state concurrence generated by the SSE in Eq. (\ref{Eq5}) in a stochastic dissipative two-qubit system, which quantifies entanglement between the non-Hermitian qubits and reaches its maximum for maximally entangled states. In Fig. \ref{Fig3}, we compare the time evolution of the mixed-state concurrence for different noise strengths $\gamma $, with $\Gamma_2=0.5$ held fixed. For different values of the stochastic-noise strength $\gamma$, the concurrence decreases slightly away from the second EP (e.g., $\Gamma_1=0.7 $) compared to the near-EP regime, which can be attributed to the increased dissipative rate  $\Gamma_1$ in the quantum jump dynamics, as shown in Fig. \ref{Fig3}(a-d).  Interestingly, the timescale required to reach maximal concurrence does not change significantly and still decreases with increasing $\Gamma_1$. We further note a good agreement between the mixed-state concurrence and the populations of the $\left| {01} \right\rangle $ and  $\left| {10} \right\rangle $ states for different values of $\gamma $, as shown in Figs. \ref{Fig2} and \ref{Fig3}, respectively. In particular, the timescale over which the populations become balanced closely coincides with the time at which the concurrence reaches its maximum, especially for large dissipation rates of the first qubit. 

\begin{figure}
    \centering
    \includegraphics[width = \linewidth]{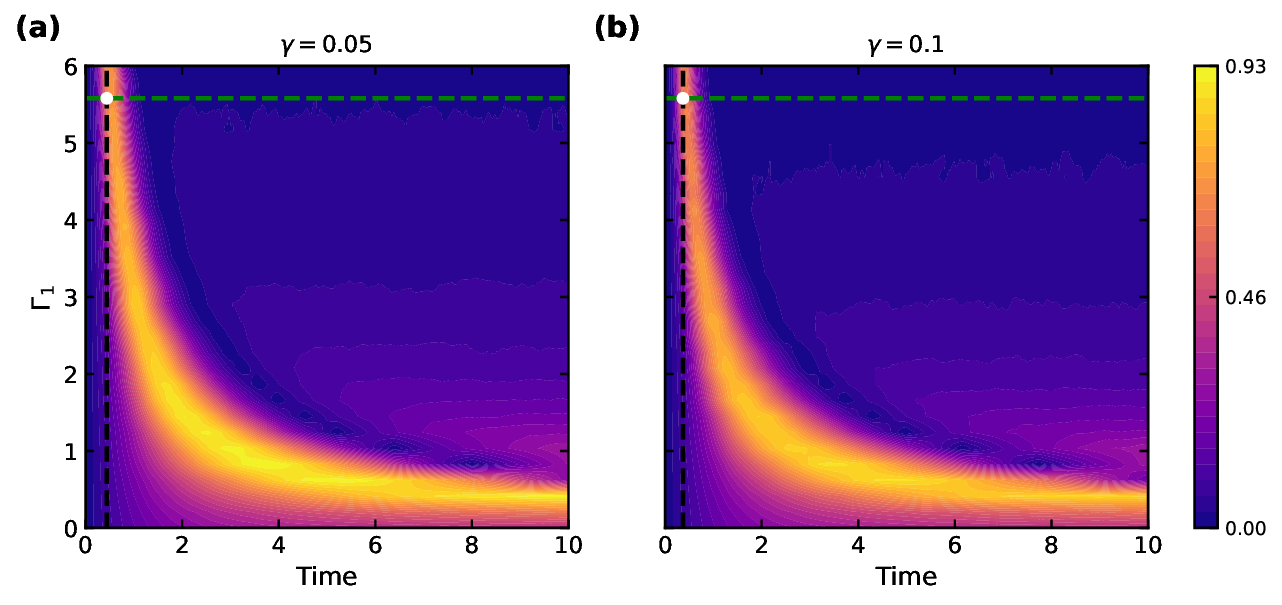}
    \caption{Density plot of the mixed-state concurrence as a function of the dissipation rate of the first qubit and time   with the dissipation rate of the second qubit ${\Gamma _2}$ fixed at 0.5 and $J=0.1$,  for different stochastic-noise strengths $\gamma$: (a) $\gamma = 0.05$ and (b) $\gamma = 0.1$. Numerical simulations are performed based on the SSE given in Eq. (\ref{Eq5}), with the system initially prepared in state $\left| {10} \right\rangle $. The results are averaged over $500$ independent single trajectories. The white point indicates a high degree of entanglement for large $\Gamma_1$.
    }
    \label{Fig4}
\end{figure}

\emph{Stochastic noise and non-Hermiticity-enhanced far from EPs---}
To gain a deeper understanding of the enhanced timescale of entanglement generation and the associated concurrence in a stochastic–dissipative two-qubit system, we analyze the  concurrence as a function of time and dissipative rate of the first qubit (Fig. \ref{Fig4}). Figure~\ref{Fig4} shows the density plot of the mixed-state concurrence as a function of $\Gamma_1$ and time across distinct $\mathcal{PT}$-symmetry regimes, for different values of  $\gamma $. The two-qubit system is initially prepared in the state $\ket{10}$, and similar results are obtained for the initial state $\ket{01}$.  From Fig. \ref{Fig4}, it is evident that the timescale corresponding to the maximum concurrence for  large $\Gamma_1$ in the $\mathcal{PT}$-symmetry-broken regime is significantly shorter than the timescale required for the system to reach the second EP and lies closer to the $\mathcal{PT}$-symmetry-unbroken region  (e.g., $0.3 \prec {\Gamma _1} \prec 0.7$) around the first EP. This indicates that entanglement generation can be both accelerated and enhanced by moving away from EPs within a symmetry-broken regime, with the effect governed by the degree of non-Hermiticity.

Furthermore, we plot the mixed‑state concurrence for different values of  $\gamma$ at fixed $\Gamma_2 = 0.5$ (Fig.~\ref{Fig4}). The characteristic timescale decreases rapidly with increasing $\Gamma_1$, while the maximum concurrence exhibits a slight reduction in the regime $\Gamma_1 > \Gamma_2$ (see Appendix \ref{app4} for further details on the ${\Gamma _1} \prec {\Gamma _2}$ regime). These results demonstrate the robustness of two‑qubit entanglement generation against stochastic noise over fast timescales, despite the reduction in the maximum concurrence. The high degree of entanglement at large ${\Gamma _1}$ in Fig. \ref{Fig4} correspond to the white point, the intersections of the green dashed and black dashed curves at the optimal entanglement parameters. Similar stochastic noise strengths have also been reported experimentally using single-photon interferometry \cite{Gao}. 

\emph{Fidelity in a stochastic dissipative two-Qubit system---}
To verify the performance of the fidelity during  entanglement generation between two non-Hermitian qubits, we characterize the distinguishability of the maximally entangled state in the presence and absence of stochastic noise using the mixed-state fidelity, known as the Uhlmann fidelity \cite{38,39,Lewalle}, 
which is  given by $F\left( {{\rho _t},\rho } \right) = {\left( {Tr\sqrt {\sqrt {{\rho _t}} \rho \sqrt {{\rho _t}} } } \right)^2}$. Here ${\rho _t}$  is the target final state obtained by acting on the qubit system initially prepared in the state $\left| {10} \right\rangle$ and  $\rho$ is the final state of the system, obtained by numerically solving the SSE in Eq. (\ref{Eq5}). 

\begin{figure}
	\centering
	\includegraphics[width =\linewidth]{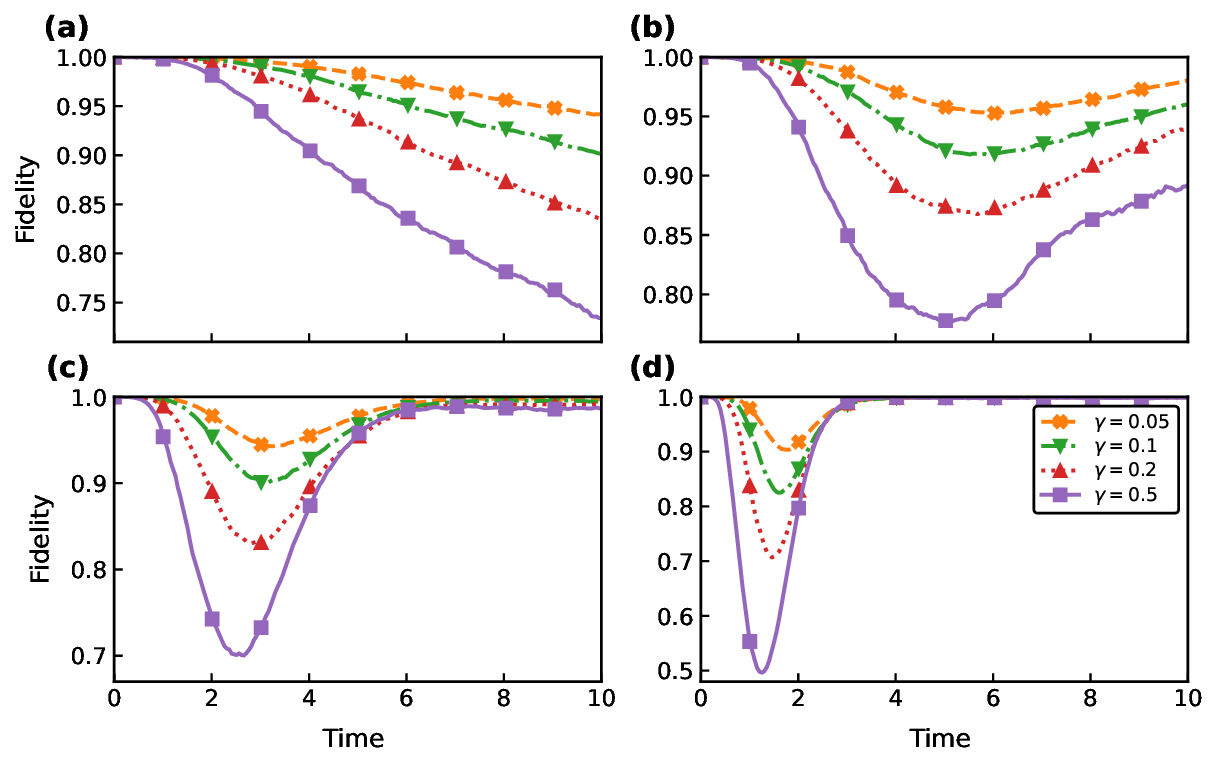}
	\caption{Fidelity evolution between states evolved
		without stochastic noise (without quantum jumps) and with
		stochastic noise, for different noise strengths $\gamma $, with $J=0.1$, ${\Gamma _2} = 0.5$ and  (a) $\Gamma_1 = 0.5$,  (b) $\Gamma_1 = 0.7$,  (c) $\Gamma_1 = 1.1$,  (d)  $\Gamma_1 = 2$. Calculations use the SSE in Eq. (\ref{Eq5}), initial state $\left| {10} \right\rangle $, and are averaged over 1000 independent single trajectories.}
	\label{Fig5}
\end{figure}

Figure \ref{Fig5} illustrates the time evolution of the fidelity under strong stochastic noise at a fixed value of ${\Gamma _2} = 0.5$ for ${\Gamma _1} = 0.5$, $0.7$,   $1.1$, and $2$, corresponding to Figs. \ref{Fig5}(a-d).  These values demonstrate that the three different $\mathcal{PT}$-symmetric phases are verified: (i) For ${\Gamma _1} = 0.5$, the system is in the $\mathcal{PT}$-unbroken phase and the fidelity in Fig. \ref{Fig5}(a) relative to the maximally entangled state decreases  with increasing evolution time, reaching a final value of approximately $0.73$ for ${\gamma } = 0.5$ exactly when the system is maximally entangled, whereas higher final fidelities are observed for other values of ${\gamma }$. (ii) For ${\Gamma _1} = 0.7$, the system lies exactly at the second EP, and the fidelity in Fig. \ref{Fig5}(b) displays an initial decay followed by a minimum at maximal entanglement  and a partial recovery, ultimately attaining a value of approximately  $0.89$ for ${\gamma } = 0.5$. (iii) For ${\Gamma _1} = 1.1$ and $2$, the system lies far from the EPs and enters the $\mathcal{PT}$-symmetry-broken phase. The fidelity in Figs. \ref{Fig5}(c) and \ref{Fig5}(d) exhibit a nonmonotonic temporal evolution, characterized by an initial decay followed by a recovery toward unity (approximately $0.99$), where the maximally entangled state does not form. At maximal entanglement, corresponding to the equality of the $\left| {01} \right\rangle $ and  $\left| {10} \right\rangle $ populations in Fig. \ref{Fig2}(c,d), the fidelity decreases with increasing stochastic-noise strength.

 Compared with Fig. \ref{Fig3}, there is a trade-off relation between the concurrence and fidelity: the maximally entangled state (highest concurrence) corresponds to lower fidelity, indicating that a high degree of
entanglement combined with noise reduces fidelity. However, for weak stochastic noise, the fidelity remains high. For $\gamma = 0.05$, the fidelity is about $0.94$, corresponding to a concurrence of approximately $0.88$ [see Figs. \ref{Fig3}(d) and \ref{Fig5}(d)]. 

\emph{Experimental outlook---} The proposed entanglement dynamics is readily realizable in a variety of experimental platforms, including two non-Hermitian superconducting qubits \cite{40,41,42}, with shared resonators mediating nearest-neighbor coupling and each qubit coupled to a readout resonator \cite{43,Zhang}. Notably, the results presented in this study remain valid even when the Hamiltonian parameters are not identical, indicating that the proposed approach should be experimentally feasible using a superconducting qubit processor as well as in  photonic systems based on the degrees of freedom of single photons \cite{Gao, 44,45, Qu, 47}. Here, we detail a single-photon experimental implementation, outlining the setup and the associated calculations.  Direct simulation of stochastic open quantum system dynamics via the SME in Eq.~(\ref{Eq4}) is experimentally challenging on photonic platforms. Alternatively, the dynamics can be reproduced using the SSE in Eq. (\ref{Eq5}), which yields results identical to those of the SME. The SSE in Eq. (\ref{Eq5}) can be rewritten as
 \begin{equation} \label{Eq6}
 i\frac{{d\left| {\psi \left( t \right)} \right\rangle }}{{dt}} = {H_\xi }\left( t \right)\left| {\psi \left( t \right)} \right\rangle, 
 \end{equation}
 where ${H_\xi }\left( t \right) = H + \left( {{i \mathord{\left/
 			{\vphantom {i 2}} \right.
 			\kern-\nulldelimiterspace} 2}} \right)L_{\phi j}^ + {L_{\phi j}} - i{\xi _j}\left( t \right){L_{\phi j}}$ and  
 the system density matrix can then be reconstructed by ensemble averaging over the corresponding stochastic state trajectories $\rho  = \left( {{1 \mathord{\left/
 			{\vphantom {1 n}} \right.
 			\kern-\nulldelimiterspace} n}} \right)\sum\nolimits_{k = 1}^n {\left| {\psi \left( t \right)} \right\rangle \left\langle {\psi \left( t \right)} \right|}$.

\begin{figure}
	\centering
	\includegraphics[width =\linewidth]{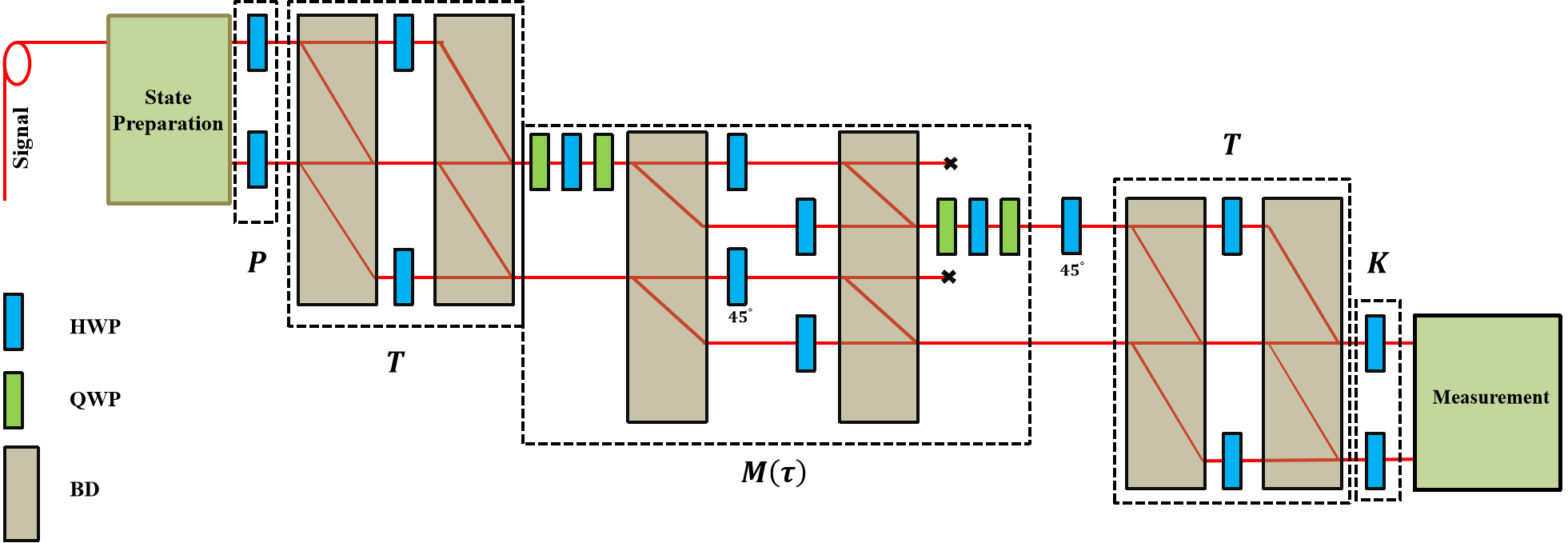}
	\caption{Schematic of the  experimental setup. The  experimental setup consists of three steps: state preparation, implementation of the evolution ${U_\xi }\left( \tau \right)$, and measurement. A photon pair is generated via spontaneous parametric down-conversion.  One photon serves as a trigger, while the other is projected into the polarization state $\left| H \right\rangle $ or $\left| V \right\rangle $. The two-qubit state is prepared by passing single photons through a polarizing beam splitter (PBS) and a beam displacer (BD); the non-unitary evolution ${U_\xi }\left( \tau \right)$ is then realized by an interferometric network consisting of beam displacers (BDs), half-wave plates (HWPs), and quarter-wave plates (QWPs), which implements the operators $T$, ${U_L}\left( \tau  \right)$, $K$, and $P$.   Finally, measurement outcomes are obtained by detecting photons.}
	\label{Fig6}
\end{figure}

Photon pairs are produced via spontaneous parametric down-conversion, where one photon acts as a trigger and the other as a heralded single photon, as shown in Fig. \ref{Fig6}. The two qubits are encoded in the spatial $\left( {\left| L \right\rangle ,\left| R \right\rangle } \right)$ and polarization $\left( {\left| H \right\rangle ,\left| V \right\rangle } \right)$  degrees of freedom of the photon. The nonunitary dynamics governed by the time-dependent Hamiltonian ${ H_\xi }\left( t \right)$ can be numerically implemented up to time $\tau$ through the associated time-evolution operator ${U_\xi }\left( \tau \right)$ \cite{Gao,47}, given by 
\begin{equation} \label{Eq7}
 {U_\xi }\left( \tau \right) = \prod\limits_{k = 1}^N {{e^{ - i{H_\xi }\left( {{t_k}} \right)\delta t}}}, 
\end{equation}
where ${t_k} = \left( {k - {1 \mathord{\left/
			{\vphantom {1 2}} \right.
			\kern-\nulldelimiterspace} 2}} \right)\delta t$ and $\delta t = {\tau  \mathord{\left/
{\vphantom {\tau  N}} \right.
\kern-\nulldelimiterspace} N}$, where $N$ is a sufficiently large natural number. The nonunitary operator 
 can be implemented according to
 \begin{equation} \label{Eq8}
 	{U_\xi }\left( \tau  \right) = K.S\left( \tau  \right).P, 
 \end{equation}
where $K$ and $P$ are controlled two-qubit gates \cite{45,46}, defined by the following $4 \times 4$ matrix representations 
 \begin{equation}\label{Eq9}
 	P = \begin{pmatrix} -1 & 0 & 0 & 0 \\ 0 & 1 & 0 & 0 \\ 0 & 0 & 0 & 1 \\ 0 & 0 & 1 & 0 \end{pmatrix}, \quad K = \begin{pmatrix} 0 & 1 & 0 & 0 \\ 1 & 0 & 0 & 0 \\ 0 & 0 & 1 & 0 \\ 0 & 0 & 0 & -1 \end{pmatrix},
 \end{equation}
and can  be realized  using  two half-wave plates (HWPs), while $S$ can  be decomposed as $S\left( \tau  \right) = T.M\left( \tau  \right).T$ with 
\begin{equation}\label{Eq10}
T = \left( {\begin{array}{*{20}{c}}
		0&1&0&0\\
		0&0&0&1\\
		1&0&0&0\\
		0&0&1&0
\end{array}} \right).
\end{equation}
This operator can be realized using two HWPs and two beam displacers (BDs) \cite{48}, while $M$  is  given  by  
\begin{equation}\label{Eq11}
	M\left( \tau  \right) = \left| L \right\rangle \left\langle L \right| \otimes {U_L}\left( \tau  \right) + \left| R \right\rangle \left\langle R \right| \otimes V\left( \tau  \right),
\end{equation}
where
\begin{equation}\label{Eq12}
	{U_L}\left( \tau  \right) = {e^{ - i{H_1}\tau }},{H_1} = \left( {\begin{array}{*{20}{c}}
			{ - i{\Gamma _1}}&J\\
			J&{ - i{\Gamma _2}}
	\end{array}} \right),
\end{equation}
and
\begin{equation}\label{Eq13}
	V\left( \tau  \right) =  - \left( {\begin{array}{*{20}{c}}
			0&1\\
			{{e^{ - \left( {{\Gamma _1} + {\Gamma _2}} \right)\tau }}}&0
	\end{array}} \right).
\end{equation}
The experimental simulation of the non-unitary operator $M\left( \tau  \right)$ can be implemented using methods similar to those of Refs. \cite{Qu,47}.  The operator $U\left( \tau  \right)$  can  be  further  decomposed according  to  ${U_L}(\tau ) = R\left( {{\theta _2},{\varphi _2},{\alpha _2}} \right)L\left( {{\theta _V},{\theta _H}} \right)R\left( {{\theta _1},{\varphi _1},{\alpha _1}} \right)$ \cite{Gao,49,50}, where the rotation operator $R\left( {{\theta _i},{\varphi _i},{\alpha _i}} \right)$($i=1,2$) can be  realized by two quarter-wave plates (QWPs) [QWP$\left( {{\theta _i}} \right)$, QWP$\left( {{\alpha _i}} \right)$], together with an HWP$\left( {{\varphi _i}} \right)$. The QWPs and HWPs are described by the matrices
\begin{equation} \notag
	Q(\theta_i)=
	\begin{pmatrix}
		\cos^2\theta_i+i\sin^2\theta_i & (1-i)\cos\theta_i\sin\theta_i\\
		(1-i)\cos\theta_i\sin\theta_i & i\cos^2\theta_i+\sin^2\theta_i
	\end{pmatrix},
\end{equation}
and
\begin{equation} \notag
	H(\varphi_i)=
	\begin{pmatrix}
		\cos(2\varphi_i) & \sin(2\varphi_i)\\
		\sin(2\varphi_i) & -\cos(2\varphi_i)
	\end{pmatrix},
\end{equation}
where $\theta_i$ and $\varphi_i$ denote the rotation angles of the QWP and HWP, respectively.
The polarization-dependent
loss operator $L\left( {{\theta _V},{\theta _H}} \right) = \left( {\begin{array}{*{20}{c}}
		0&{\sin 2{\theta _V}}\\
		{\sin 2{\theta _H}}&0
\end{array}} \right)$ can be achieved using two  beam displacers (BDs) and HWPs at ${\theta _H}$ and ${\theta _V}$. The parameters of the wave plates are chosen based on numerical calculations of the nonunitary operator given in Eq. (\ref{Eq7}). In addition, the operator $V\left( \tau  \right)$ can be realized as the polarization-dependent loss operator $L\left( {{{45}^ \circ },{\theta _H}} \right)$, or can be treated as the identity operator, since it does not affect the result  and only acts on the $\left| {00} \right\rangle $ and $\left| {11} \right\rangle $ states.

\emph{Conclusions---} 
We have studied the entanglement dynamics of two non-Hermitian qubits far from EPs, governed by a non-Hermitian Hamiltonian with stochastic noise in the dissipative rates. This entanglement  is experimentally accessible and  robust against stochastic noise. It can be generated at a speed significantly faster than that observed near EPs. These findings  demonstrate a concrete route to noise-and non-Hermiticity-enhanced entanglement generation, in which environmental decoherence becomes an asset for the development of potentially fault-tolerant quantum computing. Our results generalize to multi-qubit systems, with qubit-number–independent timescales and dynamics insensitive to proximity to EPs.

\textit{Acknowledgements}---
This work has been supported by the National Key R$\&$D
Program of China (Grant No. 2023YFA1406701). We thank members of Peng Xue’s group for helpful discussions. The open-source packages used for numerical simulations include QuTiP \cite{56,57,58}.

\textit{Data availability}---
 The data are
available from the authors upon reasonable request.

\appendix
\twocolumngrid
\appendix 

\section{ Multi-qubit entanglement with stochastic non-Hermitian  Hamiltonian }\label{app3}
\begin{figure}[H]
	\centering
	\includegraphics[width=\linewidth]{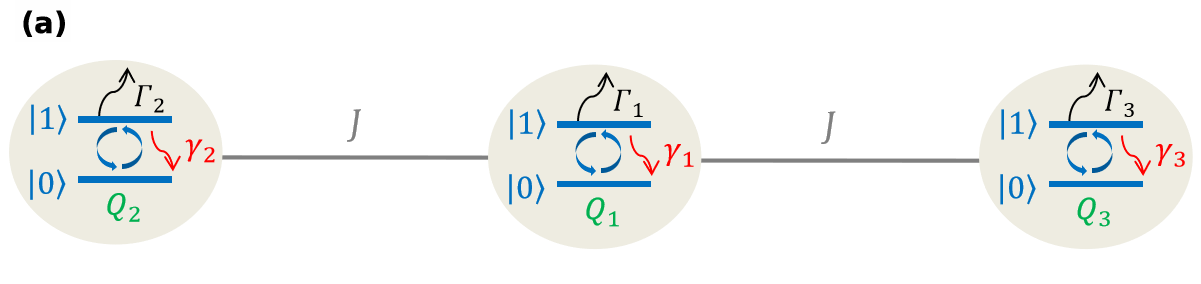}
	\hfill
	\includegraphics[width=\linewidth]{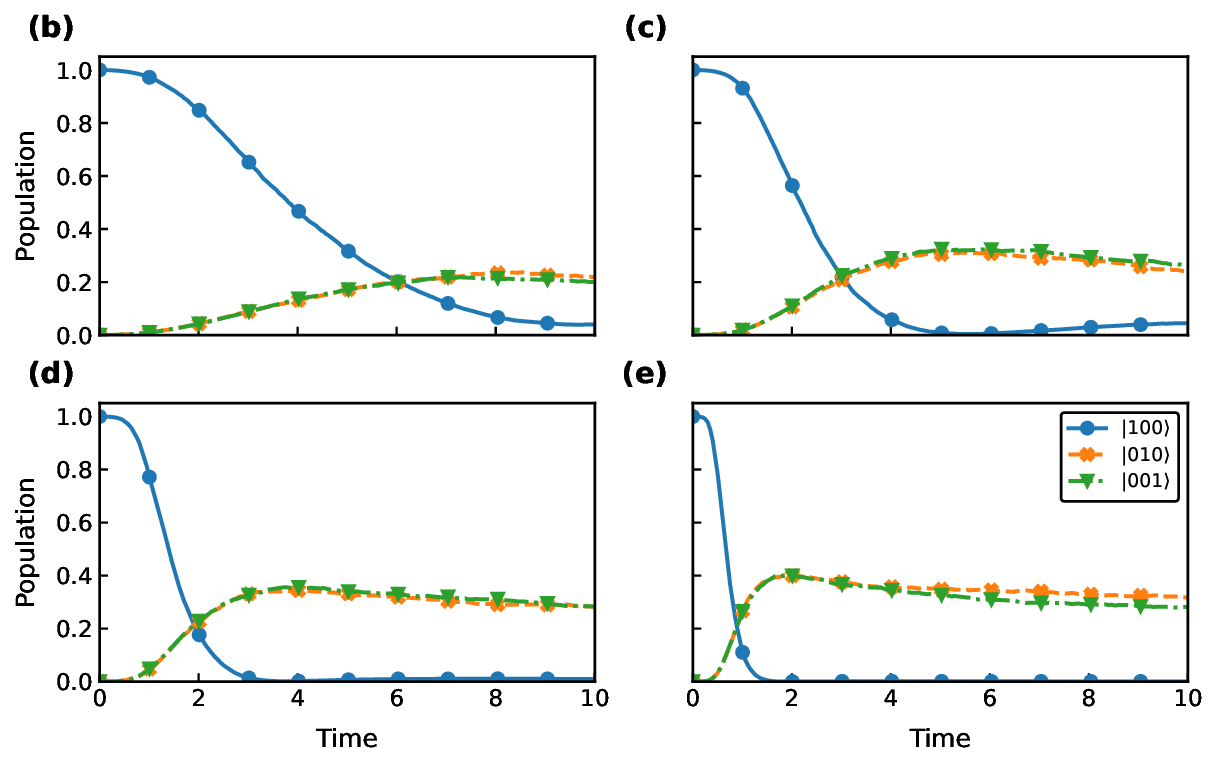}
	\caption{(a) Schematic of three non-Hermitian qubits  ${Q_1}$, ${Q_2}$ and ${Q_3}$,  where ${Q_1}$ is coupled to the other two qubits with identical coupling strength $J$. The upper level of each qubit has a dissipation rate ${\Gamma _j}\left( {j = 1, 2, 3} \right)$ and a decay rate ${\gamma _j}\left( {j = 1, 2, 3} \right)$. Populations of the three-qubit basis states as a function of time for different values of the stochastic-noise strength  and different values of the dissipative rate of the first qubit,  $\Gamma_1$: (a) $\Gamma_1 = 0.5$, (b) $\Gamma_1 = 0.78$, (c) $\Gamma_1 = 1.1$, and (d) $\Gamma_1 = 2$. The calculations are performed using the SSE in Eq. (\ref{Eq5}) for $j=1$, $2$, $3$, and the results are averaged over 1500 independent single trajectories. The initial state is given by $\left| {100} \right\rangle $, and the dissipation rates of the other qubits are fixed at  $\Gamma_2 = \Gamma_3=0.5$  for all calculations. }
	\label{Fig7}
\end{figure}

This appendix discusses the three and four qubits entanglement dynamics generated by a stochastic non‑Hermitian Hamiltonian away from exceptional points. We   consider three qubits   in a linear    array, with the  first  qubit   coupled to the second and  third qubits   with the same
coupling strength $J$ \cite{8}, described by a non-Hermitian Hamiltonian
with stochastic noise in the dissipative rates,  as shown in  Fig. \ref{Fig7}(a). The three-qubit system is described by a stochastic non-Hermitian Hamiltonian
\begin{align}  \notag 
	\tilde H =& J\left[ {\sigma _1^ + \left( {\sigma _2^ -  + \sigma _3^ - } \right) + \sigma _1^ - \left( {\sigma _2^ +  + \sigma _3^ + } \right)} \right] \\
	\label{Eq14}
	&- i\sum\limits_{j = 1}^3 {\left( {1 + \sqrt {2{\gamma _j}} {\xi _j}\left( t \right)} \right)} {L_j}.
\end{align}
 
The three-qubit dynamics in a stochastic open quantum system can be obtained by using the Hamiltonian given in Eq. (\ref{Eq14}) to solve the SSE in Eq. (\ref{Eq5}) for $j=1,2,3$,  assuming ${\gamma _1} = {\gamma _2} = {\gamma _3} \equiv \gamma $ for all  calculations. We characterize the populations of the three-qubit basis states as a function of evolution time for different values of ${\Gamma _1}$, while keeping ${\Gamma _2} = {\Gamma _3} = 0.5$, in the  presence of  the  stochastic  noise $\gamma  = 0.5$, as shown in Fig. \ref{Fig7}(b-e). During the evolution, the population is dynamically exchanged among the $\left| {001} \right\rangle $,  $\left| {010} \right\rangle $, and $\left| {100} \right\rangle $ states, and their populations equalize at a specific interaction time. The remaining basis states have negligibly small populations throughout the evolution and are therefore omitted from the plots. The equality of the three populations indicates the formation of the $W$-type maximally entangled state ${{\left( {\left| {001} \right\rangle  + \left| {010} \right\rangle  + \left| {100} \right\rangle } \right)} \mathord{\left/
		{\vphantom {{\left( {\left| {001} \right\rangle  + \left| {010} \right\rangle  + \left| {100} \right\rangle } \right)} {\sqrt 3 }}} \right.
		\kern-\nulldelimiterspace} {\sqrt 3 }}$.   It is evident from Figs. \ref{Fig7}(b-e) that the interaction time decreases as $\Gamma_1$ increases, particularly when moving away from the EPs, $\Gamma_1=0.78$ [Fig. \ref{Fig7}(c)].
Furthermore, it is noteworthy that the interaction time for the three-qubit system (Fig. \ref{Fig7}) is approximately equal to that of the two-qubit system (Fig. \ref{Fig2}) for a strong stochastic-noise strength $\gamma=0.5$, demonstrating that the proposed approach maintains its efficiency as the number of qubits increases. 
\begin{figure}
	\centering
	\includegraphics[width =\linewidth]{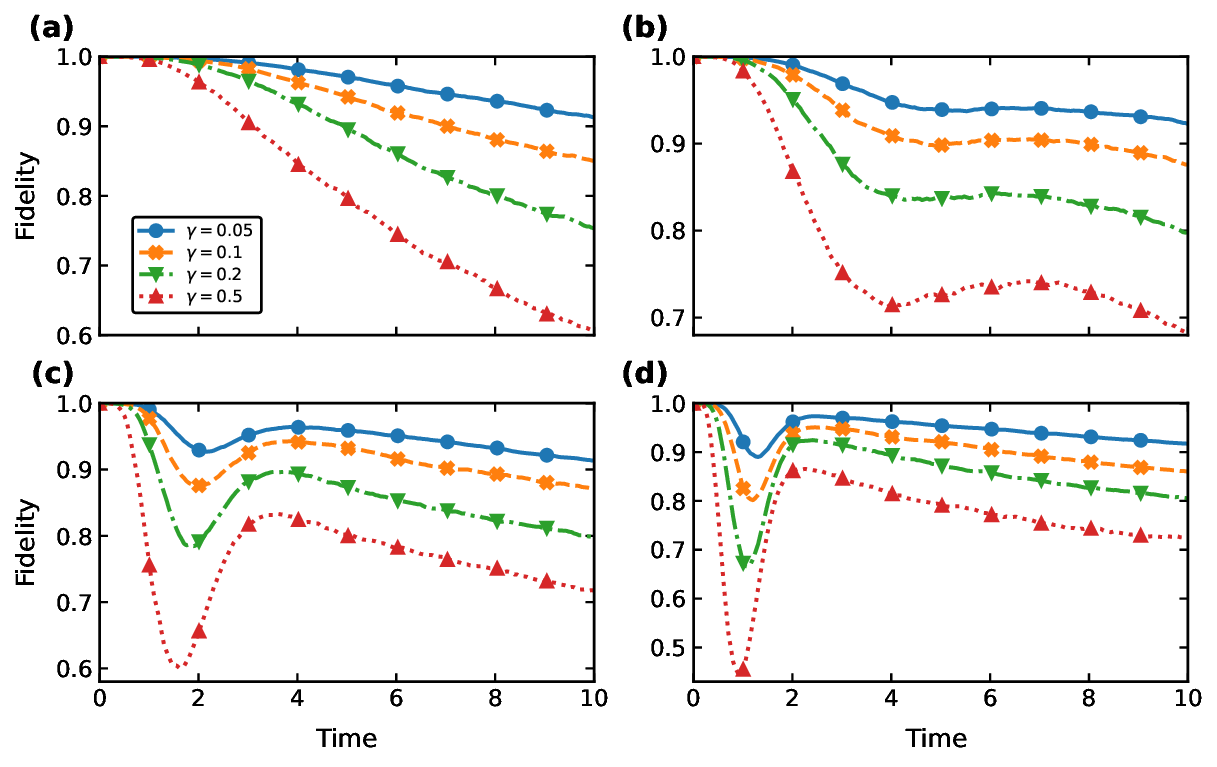}
	\caption{Time evolution of the fidelity between states evolved without stochastic noise (without quantum jumps) and with stochastic noise, for various noise strengths $\gamma $. The parameters are  ${\Gamma _2} = {\Gamma _3}= 0.5$, $J=0.1$ and  $\Gamma_1 = 0.5$ (a),  $\Gamma_1 = 0.78$ (b),  $\Gamma_1 = 1.5$ (c), $\Gamma_1 = 2.5$ (d) . Calculations are performed using the SSE in Eq. (\ref{Eq5}) for $j=1$, $2$, $3$, with initial state $\left| {100} \right\rangle $, and are averaged over 1500 independent single trajectories.}
	\label{Fig8}
\end{figure}

In addition, we also verify the performance during entanglement generation of three non‑Hermitian qubits by analyzing the mixed-state fidelity as a function of evolution time for various values of the stochastic-noise strength with different values of $\Gamma_1$ (see Fig. \ref{Fig8}). The three-qubit system is initially prepared in the state $\left| {100} \right\rangle $, and the results can also be validated for the state $\left| {001} \right\rangle $ for  all plot. As illustrated in Figs. \ref{Fig8}(c–d), the fidelity decreases with increasing stochastic noise in highly entangled states, corresponding to equal populations of the $\left| {001} \right\rangle $, $\left| {010} \right\rangle $, and $\left| {100} \right\rangle $ states in Fig. \ref{Fig7}(b–e). However, for small values of the stochastic noise parameter $\gamma$, the fidelity remains high and approaches unity elsewhere, exhibiting behavior similar to that observed in the two-qubit system (Fig. \ref{Fig3}).
\begin{figure}
	\centering
	\includegraphics[width=\linewidth]{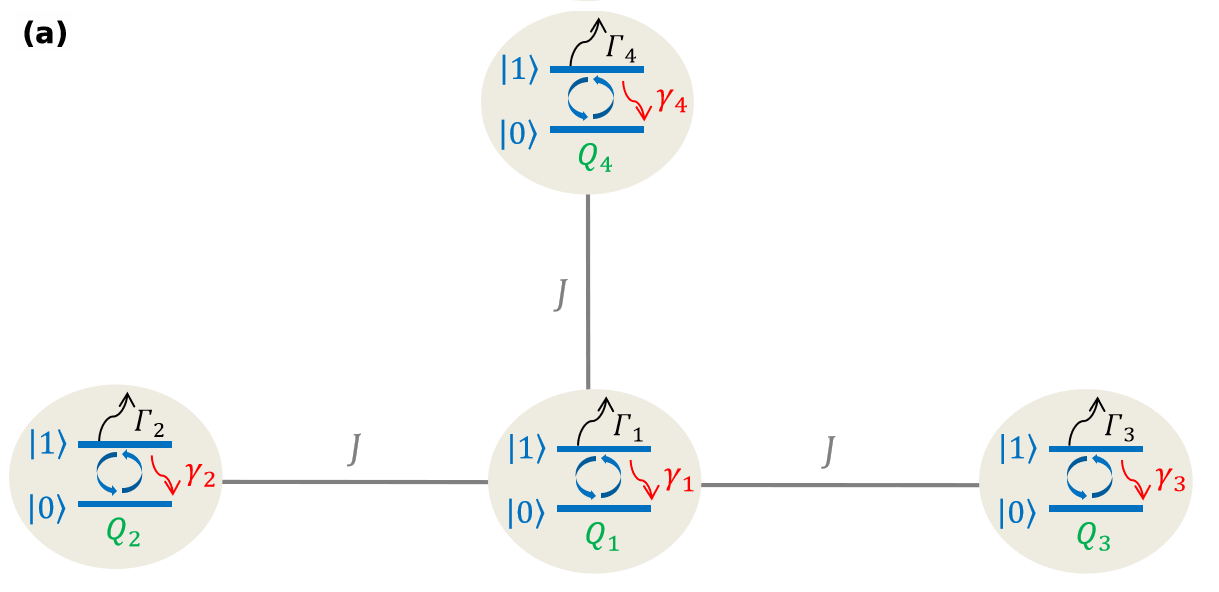}
	\hfill
	\includegraphics[width=\linewidth]{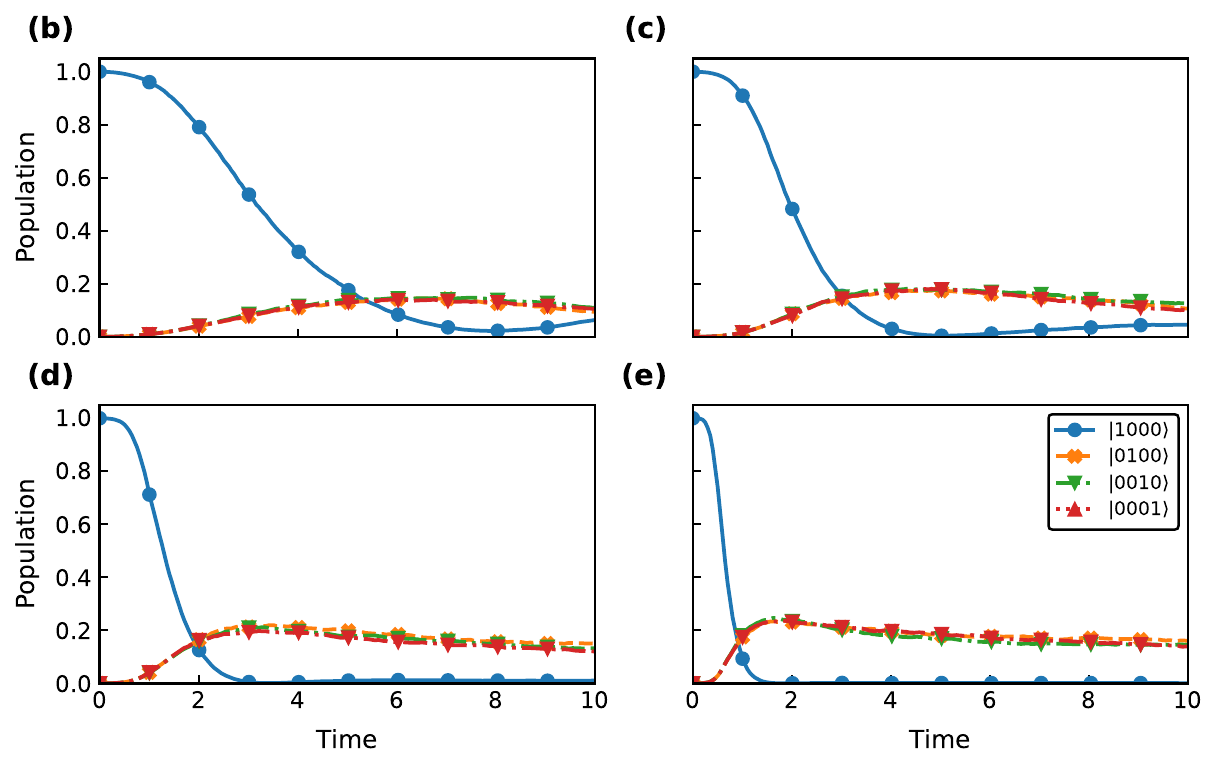}
	\caption{(a) Schematic of four non-Hermitian qubits  ${Q_1}$, ${Q_2}$, ${Q_3}$ and ${Q_4}$,  where ${Q_1}$ is coupled to the other three qubits with identical coupling strength $J$. The upper level of each qubit has a dissipation rate ${\Gamma _j}\left( {j = 1, 2, 3, 4} \right)$ and a decay rate ${\gamma _j}\left( {j = 1, 2, 3, 4} \right)$. Populations of the Four-qubit basis states as a function of time for different values of the stochastic-noise strength and and different values of the dissipative rate of the first qubit,  $\Gamma_1$: (a) $\Gamma_1 = 0.5$, (b) $\Gamma_1 = 0.78$, (c) $\Gamma_1 = 1.1$, and (d) $\Gamma_1 = 2$. The calculations are performed using the SSE in Eq. (\ref{Eq5}) for $j=1$ ,$2$, $3$, $4$, and the results are averaged over 1500 independent single trajectories. The initial state is given by $\left| {1000} \right\rangle $, and the dissipation rates of the other qubits are fixed at  $\Gamma_2 = \Gamma_3= \Gamma_4= 0.5$. }
	\label{Fig9}
\end{figure}

Importantly, we note that the proposed scheme can be naturally extended to multi-qubit systems by coupling a single qubit to multiple others. As a concrete example, we consider a non-Hermitian four-qubit system in which the first qubit is uniformly coupled to the remaining three qubits with identical coupling strength J, while the dissipative rates of all qubits are subject to stochastic noise, as illustrated in Fig. \ref{Fig9}(a). The dynamics of the multi-qubit system are governed by the non-Hermitian Hamiltonian 
\begin{align}  \label{Eq15} 
	\tilde H = J\left( {\sum\limits_{j = 2}^n {\sigma _1^ + } \sigma _j^ -  + H.c} \right)
	 - i\sum\limits_{j = 1}^n {\left( {1 + \sqrt {2{\gamma _j}} {\xi _j}\left( t \right)} \right)} {L_j}.
\end{align}

The evolution of the populations corresponding to the $\left| {1000} \right\rangle ,\left| {0100} \right\rangle ,\left| {0010} \right\rangle$ and $\left| {0001} \right\rangle$ states under stochastic noise, obtained using the Hamiltonian given in Eq. (A2)  ($n = 4$) to solve Eq. (\ref{Eq5}) for $j = 1,2,3,4$, is shown in Fig. \ref{Fig9}(b–e). We assume that  $\gamma_1 = \gamma_2 = \gamma_3 = \gamma_4 \equiv  \gamma$ throughout the calculations, and the four-qubit system is initially prepared in the state $\left| {1000} \right\rangle$. It is clear from Fig. \ref{Fig9} that all populations become equal at a specific interaction time, which decreases with increasing $\Gamma_1$, while  $\Gamma_2 = \Gamma_3 = \Gamma_4 =0.5$  is held fixed. This indicates that the four-qubit system realizes a W-type maximally entangled state $\left| W \right\rangle  = {{\left( {\left| {1000} \right\rangle  + \left| {0100} \right\rangle  + \left| {0010} \right\rangle  + \left| {0001} \right\rangle } \right)} \mathord{\left/
		{\vphantom {{\left( {\left| {1000} \right\rangle  + \left| {0100} \right\rangle  + \left| {0010} \right\rangle  + \left| {0001} \right\rangle } \right)} 2}} \right.
		\kern-\nulldelimiterspace} 2}$. We further note that the interaction timescale matches that observed in the two‑qubit system  (Fig. \ref{Fig2} for $\gamma = 0.5$)  and the three‑qubit system [Fig. \ref{Fig7}(b-e)]. This reveals an additional robustness of multipartite entanglement dynamics: the relevant timescale is independent of the number of qubits, and entanglement generation remains resilient against environmental decoherence. These features enhance the experimental feasibility of the proposal.
	
	\section{ Impact of a deterministic decay channel in the SME  }\label{app4}
	
\begin{figure}
		\centering
		\includegraphics[width = \linewidth]{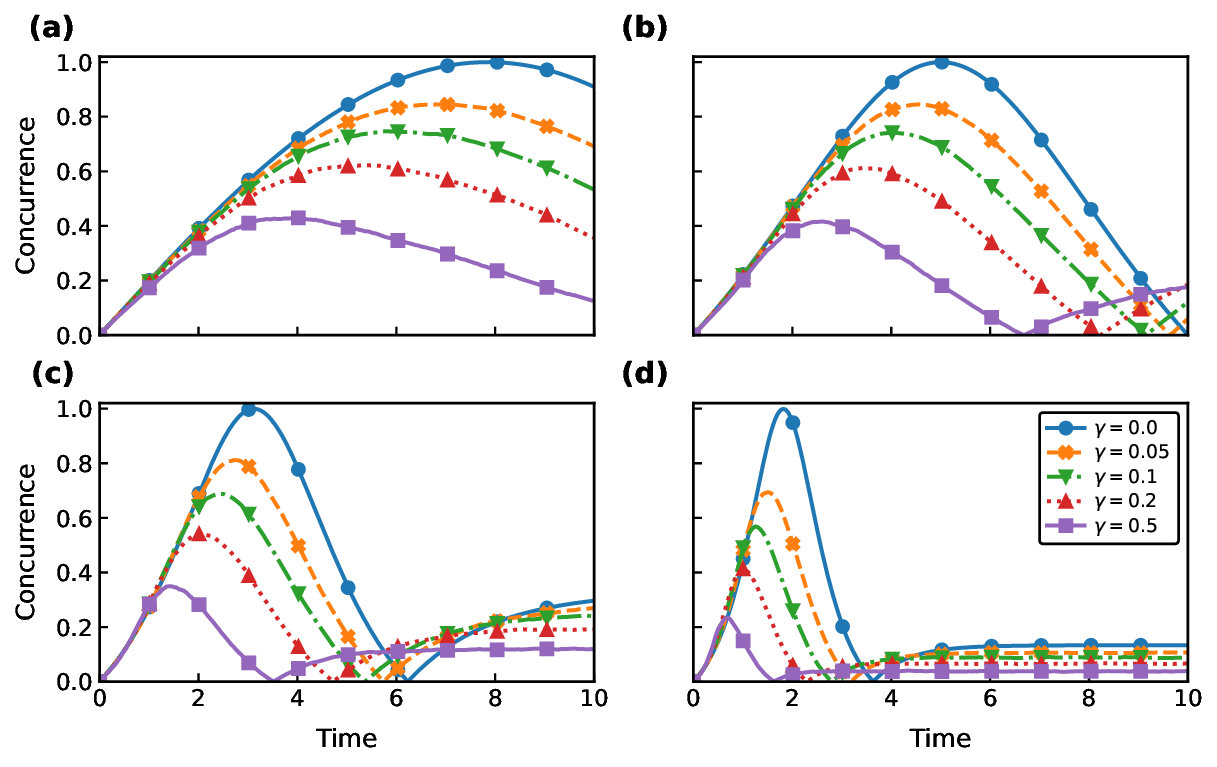}
		\caption{Mixed-state  concurrence as function  of evolution time for  different  stochastic noise  strenghs $\gamma$, with  ${\Gamma _2} = 0.5$, $J=0.1$ and for different values of ${\Gamma _1}$: (a) $\Gamma_1 = 0.5$, (b) $\Gamma_1 = 0.7$, (c) $\Gamma_1 = 1.1$, and (d) $\Gamma_1 = 2$. Calculations are performed based on the SME given in Eq. (\ref{EqB1}), with the system initialized in the state $\left| {10} \right\rangle $  and results averaged over 1500 independent single trajectories.  }
		\label{Fig10}
	\end{figure}

	In the main text, the primary decay mechanism considered is that associated with the stochastic term in the SME [Eq. (\ref{Eq3})] and the SSE [Eq. (\ref{Eq5})]. However, to assess whether our results are observable in realistic experimental setups, a decay channel in the deterministic part must also be included. The second term captures this effect through the double anticommutator in the original SME, 
	\begin{align}\notag 
		d\rho  =&  { - i\left( {H\rho  - \rho {H^ + }} \right)dt + \frac{1}{2}\left\{ {{L_{\phi j}},\left\{ {{L_{\phi j}},\rho } \right\}} \right\}} dt \\
		\label{EqB1}
		&- \left\{ {{L_{\phi j}},\rho } \right\}d{W_j}\left( t \right).
	\end{align}
	Below, considering both effects related to the stochastic and deterministic terms—characterized by the dephasing rate ${\gamma _{\phi j}}$ and identical quantum jumps. These effects arise from an anti-dephasing master equation that enriches the manifold of possible steady states, as discussed in Ref. \cite{32, 33}. We assume that the two non-Hermitian qubits have the same decay rate for their excited states, i.e., $\gamma_1 = \gamma_2 \equiv  \gamma$. The concurrence evolution is shown for different values of the dissipation rate of the first qubit, $\Gamma_1=0.5$ [Fig. \ref{Fig10}(a)], $0.7$ [Fig. \ref{Fig10}(b)], $1.1$ [Fig. \ref{Fig10}(c)], and $2$ [Fig. \ref{Fig10}(d)], for various values of $\gamma$, with the initial state $\left| {10} \right\rangle $. The parameters are the same as those in Fig. \ref{Fig3} of the main text. Figure \ref{Fig10} shows that the behavior of the concurrence is similar to that in Fig. \ref{Fig3} of the main text: the timescale decreases with increasing ${\Gamma _1}$ accompanied by a slight reduction in the maximal concurrence. We find that including this effect  modifies the maximal concurrence for each value of $\gamma$; however, it does not increase the timescale. Thus, the mechanism of entanglement generation under stochastic noise remains robust against decoherence, making it promising for experimental realization with current quantum technologies.

\begin{figure}[H]
		\centering
		\includegraphics[width = \linewidth]{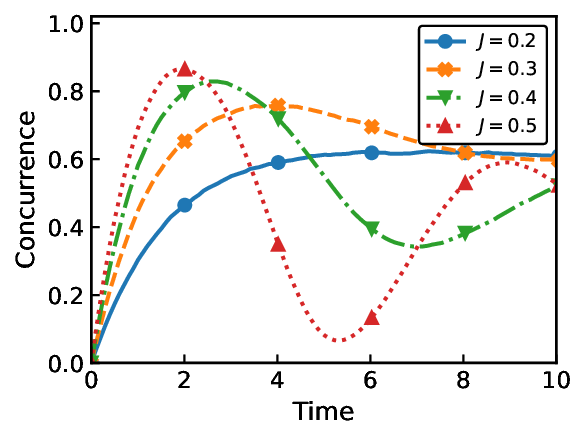}
		\caption{Mixed-state  concurrence as function  of evolution time for  different  coupling strengths $J$, with  ${\Gamma _2} = 0.5$, $\gamma=0.2$. Calculations are performed based on the SME given in Eq. (\ref{EqB1}), with the system initialized in the state $\left| {10} \right\rangle $  and results averaged over 1500 independent single trajectories.  }
		\label{Fig11}
	\end{figure}

	We also investigate the effect of the dissipation rate $\Gamma_1$ in the regime where $\Gamma_1 < \Gamma_2 = 0.5$, compared with the  ${\Gamma _1} \succ {\Gamma _2}$ regime discussed in the main text. Figure \ref{Fig11} shows the evolution of concurrence for ${\Gamma _1}= 0.1$, which is smaller than ${\Gamma _2}$. This dissipation rate lies away from the EPs. We note that a maximally entangled state can be obtained by tuning the coupling strength to  $J= 0. 4$,   $0.5$, with a maximum concurrence exceeding that observed in   the  ${\Gamma _1} \succ {\Gamma _2}$ regime. Nonetheless, the time required to attain the maximum concurrency is still longer than the value reported in the main text.

Finally, we identify robust entanglement dynamics, including both stochastic and deterministic decay with dissipation rates for each qubit in the quantum jump processes, which do not alter the time at which the concurrence reaches its maximum, although it reduces this maximum value. Moreover, our results demonstrate entanglement dynamics at an optimal dissipation rate ${\Gamma _1}$ of the first qubit, ensuring that the system remains far from EPs and is robust against   noise.	 


\begin{thebibliography}{82}%
	
	\bibitem{1}Z.-Z. Li, W. Chen, M. Abbasi, K. W. Murch, and
	K. B. Whaley, Speeding Up Entanglement Generation by Proximity to Higher-Order Exceptional Points,
	Phys. Rev. Lett. 131, 100202 (2023).
	
	\bibitem{2}A. Kumar, K. W. Murch, and Y. N. Joglekar, Maximal
	quantum entanglement at exceptional points via unitary
	and thermal dynamics, Phys. Rev. A 105, 012422 (2022).
	
	\bibitem{3}C. G. Feyisa, J. You, H.-Y. Ku, and H. Jen, Accelerating
	multipartite entanglement generation in non-Hermitian
	superconducting qubits, Quantum Sci. Technol. 10, 025021(2025).
	
	\bibitem{4}C. G. Feyisa, C. Y. Liu, M. S. Hasan, J. S. You, H. Y. Ku, and H. H. Jen, Robustness of multipartite entangled states in passive PT-symmetric qubits, Phys. Rev. Research 7, 033060(2025).
	
	\bibitem{5}B. A. Tay, and Y. S. H'ng, Entanglement generation across exceptional points in a two-qubit open quantum system: The role of initial states, Phys. Rev. E 112, 024133(2025).
	
	\bibitem{Tang} Z. Tang, T. Chen, and X. Zhang, Highly efficient transfer of quantum state and robust generation of entanglement state
	around exceptional lines, Laser Photonics Rev. 2300794 (2023).
	
	\bibitem{6}Z. Tang, T. Chen, X. Tang, and X. Zhang, Topologically protected entanglement switching around exceptional points, Light Sci Appl 13, 167 (2024).
	
	\bibitem{7}Y. L. Fang,  J. L. Zhao,  D. X. Chen,  Y. H. Zhou, Y. Zhang, Q. C. Wu, C. P. Yang, and  F. Nori, Entanglement dynamics in anti-PT-symmetric systems, Phys. Rev. Research 4, 033022 (2022).
	
	\bibitem{8}X. W. Zheng,   J. C. Zheng,   X. F. Pan, L. H. Lin,  P. R. Han, and   P. B. Li, Conditional acceleration of entanglement generation enabled by dissipation, Phys. Rev. A 111, 012402(2025).
	
	\bibitem{9} C. Hotter,  A. Kosior,  H. Ritsch,  and  K. Gietka, Conditional Entanglement Amplification via Non-Hermitian Superradiant Dynamics, Phys. Rev. Lett. 134, 233601(2025).
	
	\bibitem{10}A. A. Budini, Quantum systems subject to the action of classical stochastic fields, Phys. Rev. A 64, 0521101(2001).
	
	\bibitem{11}
	A. Kiely, Exact classical noise master equations: Applications and connections, Europhys. Lett. 134, 10001 (2021).
	
	\bibitem{12} N. V. Kampen, Stochastic Processes in Physics and Chemistry, North-Holland Personal Library (Elsevier Science, 2011).
	
	\bibitem{Chenu} A. Chenu, M. Beau,
	J. Cao, and A. del Campo, Quantum simulation of generic many-body open system dynamics using
	classical noise. Phys. Rev. Lett. 118, 140403 (2017).
	
	\bibitem{13}
	A. Sander, M. Fröhlich,  M. Eigel,  J. Eisert,  P. Gelß, M. Hintermüller, R. M. Milbradt, R. Wille, and  C. B. Mendl,  Large-scale stochastic simulation of open quantum systems, Nat Commun 16, 11074 (2025).
	
	\bibitem{14}
	P. M. Azcona, A. Kundu, A. d. Campo, and
	A. Chenu, Stochastic operator variance: An observable
	to diagnose noise and scrambling, Phys. Rev. Lett. 131,
	160202 (2023).
	
	\bibitem{15} K. Mølmer, K. B. Sørensen, Y. Castin, and J. Dalibard, "A Monte Carlo wave function method in quantum optics," in Optical Society of America Annual Meeting, Technical Digest Series (Optica Publishing Group, 1992).
	
	\bibitem{16} J. Dalibard, Y. Castin, and K. Mølmer, Wave-function approach to
	dissipative processes in quantum optics. Phys. Rev. Lett. 68,
	580 (1992).
	
	\bibitem{17} H. M. Wiseman, Quantum trajectories and quantum measurement theory, Quantum Semiclass. Opt. 8, 205 (1996).
	
	\bibitem{18} R. L. Hudson, and  K. R. Parthasarathy , Quantum Ito’s formula and
	stochastic evolutions. Comm. Math. Phys. 93, 301 (1984).
	
	\bibitem{19} H. M. Wiseman and G. J. Milburn, Quantum Measurement and Control (Cambridge University Press, Cambridge, 2009).
	
	\bibitem{20} H. J. Carmichael,  An Open Systems Approach to Quantum Optics.
	(Springer-Verlag, Berlin Heidelberg, 1993).
	
	\bibitem{21} J. Wang, H. M. Wiseman, and G. J. Milburn, Dynamical creation of entanglement by homodyne-mediated feedback, Phys. Rev. A 71, 042309 (2005).
	
	\bibitem{22} M. Sarovar, H. S. Goan, T. P. Spiller, and G. J. Milburn, High-fidelity measurement and quantum feedback control in circuit QED,
	Phys. Rev. A 72, 062327 (2005).
	
	\bibitem{23} J. Kerckhoff, L. Bouten, A. Silberfarb, and H. Mabuchi, Physical model of continuous two-qubit parity measurement in a cavity-QED network,
	Phys. Rev. A 79, 024305 (2009).
	
	\bibitem{24} Z. Liu, L. Kuang, K. Hu, L. Xu, S. Wei, L. Guo, and
	X. Q. Li, Deterministic creation and stabilization of entanglement in circuit QED by homodyne-mediated feedback control, Phys. Rev. A 82, 032335 (2010).
	
	\bibitem{25} L. Martin, M. Sayrafi, and K. B. Whaley, What is the
	optimal way to prepare a Bell state using measurement
	and feedback?, Quantum Sci. Technol. 2,
	044006 (2017).
	
	\bibitem{Martin} L. Martin, F. Motzoi, H. Li, M. Sarovar, and K. B.
	Whaley, Deterministic generation of remote entanglement with active quantum feedback, Phys. Rev. A
	92, 062321 (2015).
	
	\bibitem{26} S. Zhang, L. S. Martin, and K. B. Whaley, Locally optimal measurement-based quantum feedback with application to multiqubit entanglement generation, Phys.
	Rev. A 102, 062418 (2020).
	
	\bibitem{27} J. E. Reiner, W. P. Smith, L. A. Orozco, H. M. Wiseman,
	and J. Gambetta, Quantum feedback in a weakly driven cavity QED system, Phys. Rev. A 70, 023819 (2004).
	
	\bibitem{28} C. Sayrin, I. Dotsenko, X. Zhou, B. Peaudecerf,
	T. Rybarczyk, S. Gleyzes, P. Rouchon, M. Mirrahimi,
	H. Amini, M. Brune, et al., Real-time quantum feedback prepares and stabilizes photon number states, Nature 477, 73 (2011).
	
	\bibitem{29} K. C. Cox, G. P. GReve, J. M. Weiner, and J. K. Thompson, Deterministic Squeezed States with Collective Measurements and Feedback, Phys. Rev. Lett. 116, 093602 (2016).
	
	\bibitem{30} L. S. Martin, W. P. Livingston, S. H. Gourgy,
	H. M. Wiseman, and I. Siddiqi, Implementation of a
	canonical phase measurement with quantum feedback,
	Nat. Phys. 16, 1046–1049 (2020).
	
	\bibitem{31} S. Greenfield
	,  L. Martin, F. Motzoi,
	 K. B. Whaley, J. Dressel,
	  and E. M. Levenson-Falk
	  , Stabilizing two-qubit entanglement with dynamically decoupled active feedback,
	Phys. Rev. Applied 21, 024022(2024).
	
	\bibitem{32} P. M. Azcona, A. Kundu, A. Saxena, A. d. 
	Campo, and A. Chenu, Quantum dynamics with stochastic non-hermitian hamiltonians, Phys. Rev. Lett. 135,
	010402 (2025).
	
	\bibitem{33} P. M. Azcona, M. Sarkis, A. Tkatchenko, and  A. Chenu, Magic Steady State Production:
	Non-Hermitian and Stochastic pathways (2025), arXiv:2507.08676.
	
	 \bibitem{Gal} Y. L. Gal, X. Turkeshi, and M. Schir\`o, Entanglement
	dynamics in monitored systems and the role of quantum
	jumps, PRX Quantum 5, 030329 (2024).
	
	\bibitem{34}A. Barchielli and M. Gregoratti, Quantum trajectories
	and measurements in continuous time (Springer-Verlag,
	Berlin, Heidelberg, 2009).
	
	\bibitem{35} A. N. Jordan and I. A. Siddiqi, Quantum Measurement: Theory and Practice (Cambridge University
	Press, 2024).
	
	\bibitem{36} P. Goetsch and R. Graham, Linear stochastic wave equations for continuously measured quantum systems, Phys. Rev. A 50, 5242
	(1994).
	
	\bibitem{37} W. K. Wootters, Entanglement of formation of an arbitrary state of two qubits, Phys. Rev. Lett. 80, 2245 (1998).
	
	\bibitem{Gao} H. Gao, K. Sun, D. Qu, K. Wang, L. Xiao, W. Yi,
	and P. Xue, Photonic Chiral State Transfer near the
	Liouvillian Exceptional Point, Phys. Rev. Lett. 134,
	146602 (2025).
	
	\bibitem{38} A. Uhlmann, The ``transition probability'' in the state space of a $^*$-algebra, Rep. Math. Phys. 9, 273 (1976).
	
	\bibitem{39} R. Jozsa, Fidelity for mixed quantum states, J. Mod. Opt.
	41, 2315 (1994).
	
	\bibitem{Lewalle} P. Lewalle, Y. Zhang, and K. B. Whaley, Optimal Zeno
	dragging for quantum control: A shortcut to Zeno with
	action-based scheduling optimization, PRX Quantum 5,
	020366 (2024).
	
	\bibitem{40} J. Koch, T. M. Yu, J. Gambetta, A. A. Houck, D. I.
	Schuster, J. Majer, A. Blais, M. H. Devoret, S. M. Girvin,
	and R. J. Schoelkopf, Charge insensitive qubit design derived from the Cooper pair box, Phys. Rev. A 76, 042319 (2007).
	
	\bibitem{41} M. Naghiloo, M. Abbasi, Y. N. Joglekar, and K. Murch, Quantum state tomography across the exceptional point in a single
	dissipative qubit, Nat. Phys. 15, 1232 (2019).
	
	\bibitem{42} W. Chen, M. Abbasi, Y. N. Joglekar, and K. W. Murch, Quantum jumps in the non-Hermitian dynamics of a superconducting
	qubit, Phys. Rev. Lett. 127, 140504 (2021).
	
	\bibitem{43} C. Gaikwad, D. Kowsari, C. Brame, X. Song, H. Zhang,
	M. Esposito, A. Ranadive, G. Cappelli, N. Roch, E. M.
	Levenson-Falk, and K. W. Murch, Entanglement assisted probe of the non-Markovian to Markovian transition in open quantum system dynamics, Phys. Rev. Lett. 132, 200401 (2024).
	
	\bibitem{Zhang}  H. L. Zhang, P. R. Han, F. Wu, W. Ning, Z. B. Yang,
	and S.-B. Zheng, Experimental observation of non-markovian quantum exceptional points, Phys. Rev. Lett.
	135, 230203 (2025).
	
	\bibitem{Kampen}  V. Kampen, and N. Godfried, Stochastic processes in
	physics and chemistry, Vol. 1. Elsevier (1992).
	
	\bibitem{44} P. Xue, L. Xiao, G. Ruffolo, A. Mazzari, T. Temistocles,
	M. T. Cunha, and R. Rabelo, Synchronous observation of
	Bell nonlocality and state-dependent contextuality, Phys.
	Rev. Lett. 130, 040201 (2023).
	
	\bibitem{45} K. Wang, R. Xia, L. Bresque, and P. Xue, Experimental demonstration of a quantum engine driven by entanglement and local measurements, Phys. Rev. Res. 4,
	L032042 (2022).
	
	\bibitem{46} K. Wang, Y. Shi, L. Xiao, J. Wang, Y. N. Joglekar, and P. Xue,
	Experimental realization of continuous-time quantum walks on
	directed graphs and their application in PageRank, Optica 7,
	1524 (2020).
	
	\bibitem{Qu} D. Qu, I. I. Arkhipov, H. Gao, K. Wang, L. Xiao, F. Nori, and P. Xue, Selective chiral multistate switching via the dynamic interplay
	of diabolic and exceptional points, Phys. Rev. Lett. 136, 086603 (2026).
	
	\bibitem{47} L. Xiao, D. Qu, K. Wang, H. W. Li, J. Y. Dai, B.
	D\'ora, M. Heyl, R. Moessner, W. Yi, and P. Xue, Nonhermitian kibble-zurek mechanism with tunable complexity in single-photon interferometry, PRX Quantum
	2, 020313 (2021).
	
	\bibitem{48} Y. Li, J. Xing, D. Qu, H. Gao, L. Xiao, J. M. Liu,
	Y. Xiao, and P. Xue, Temporal asymmetry in entanglement distillation, Phys. Rev. Lett. 135, 170801(2025).
	
	\bibitem{49} L. Xiao, K. Wang, X. Zhan, Z. Bian, K. Kawabata,
	M. Ueda, W. Yi and P. Xue, Observation of critical
	phenomena in parity-time-symmetric quantum dynamics, Phys. Rev.Lett. 123, 230401 (2019).
	
	\bibitem{50} H. Gao, K. K. Wang, L. Xiao, M. Nakagawa, N. Matsumoto, D. Qu, H. Lin, M. Ueda, and P. Xue, Experimental observation of the yang-lee quantum criticality in open quantum systems, Phys. Rev. Lett. 132,
	176601 (2024).
	
	\bibitem{56} J. R. Johansson, P. D. Nation, and F. Nori, QuTiP: An
	open-source Python framework for the dynamics of open
	quantum systems, Comput. Phys. Commun. 183, 1760
	(2012).
	
	\bibitem{57} J. R. Johansson, P. D. Nation, and F. Nori, QuTiP 2: A Python
	framework for the dynamics of open quantum systems,
	Comput. Phys. Commun. 184, 1234 (2013).
	
	\bibitem{58} N. Lambert, E. Giguère, P. Menczel, B. Li, P. Hopf, G.
	Suárez, M. Gali, J. Lishman, R. Gadhvi, R. Agarwal,
	A. Galicia, N. Shammah, P. Nation, J. R. Johansson, S.
	Ahmed, S. Cross, A. Pitchford, and F. Nori, QuTiP 5: The
	quantum toolbox in Python, Phys. Rep. 1153, 1 (2026). 
\end{thebibliography}
\end{document}